\documentclass[showpacs,twocolumn,floatfix,superscriptaddress,prb,eqsecnum]{revtex4}
\usepackage{graphicx}
\usepackage{amsmath}
\begin{document}
\title{Weak localization of the open kicked rotator}
%\author{A. B. C.}
\author{J. Tworzyd{\l}o}
\affiliation{Instituut-Lorentz, Universiteit Leiden, P.O. Box 9506, 2300 RA
Leiden, The Netherlands}
\affiliation{Institute of Theoretical Physics, Warsaw University, Ho\.{z}a 69,
00--681 Warsaw, Poland}
\author{A. Tajic}
\affiliation{Instituut-Lorentz, Universiteit Leiden, P.O. Box 9506, 2300 RA
Leiden, The Netherlands}
\author{C.W.J. Beenakker}
\affiliation{Instituut-Lorentz, Universiteit Leiden, P.O. Box 9506, 2300 RA
Leiden, The Netherlands}
\date{May 2004}
\begin{abstract}

We present a numerical calculation of the weak localization peak in
the magnetoconductance for a stroboscopic model of a 
chaotic quantum dot. The magnitude of the peak is close to the universal
prediction of random-matrix theory. The width depends on the classical
dynamics, but this dependence can be accounted for by a single parameter:
the level curvature around zero magnetic field of the closed system.

\end{abstract}
\maketitle

\section{Introduction}

Random-matrix theory (RMT) makes system-independent (``universal'')
predictions about quantum mechanical systems with a chaotic classical
dynamics \cite{Haa90,Bee97,Guh98,Alh00}. The presence or absence of time-reversal symmetry
(TRS) identifies two universality classes. RMT is also capable of 
describing the crossover between the universality classes, e.g.\ when 
TRS is broken by the application of a magnetic field $B$. The crossover is 
predicted to depend on a single system-specific parameter, being the mean 
absolute curvature of the energy levels $E_i$ around $B=0$.
More precisely, a universal magnetic-field dependence of spectral correlations is
predicted when $B$ is rescaled by the characteristic field 
\begin{equation}
B_c = \left( \frac{1}{\Delta}
\left <
\left| \frac{d^2 E_i}{d B^2} \right|_{B=0}
\right >
\right)^{-1/2},
\label{bc}
\end{equation}
with $\Delta$ the mean level spacing. This prediction has been tested in
a variety of computer simulations \cite{Boh95,Yan95,Shu97}.

In open systems there exists a similar prediction of universality for
transport properties, but now the characteristic field depends also on
the conductance $g$ of the point contacts that couple the
chaotic quantum dot to electron reservoirs \cite{Bar93,Plu94,Efe95,Fra95}. 
A universal magnetic field dependence is predicted if $B$ is rescaled by 
$B_c \sqrt{g}$,  provided $g$ is large compared to the conductance 
quantum $e^2/h$. To provide a numerical test of this prediction is the 
purpose of this paper.

We present a computer simulation of the open quantum kicked rotator
\cite{Oss03,Jac03,Two03}, which is a stroboscopic model of a quantum dot coupled 
to electron reservoirs by ballistic point contacts. The ensemble
averaged conductance increases upon breaking of TRS, as a manifestation of 
weak localization. The height, width, and lineshape of the weak localization
peak are compared with the predictions of RMT.

The simulation itself is straightforward, but the formulation of the model
is not. There exist several ways to break TRS in the closed kicked
rotator \cite{Izr86,Izr90,Blu92,Tha93}
and related models \cite{Car98,Dit00,Jon03,Jonck03}. 
When opening up the system one needs to ensure that the scattering matrix satisfies the
reciprocity relation 
\begin{equation}
S(-B)=S^T(B).
\label{Srec}
\end{equation} 
(The superscript $T$ indicates the transpose
of the scattering matrix $S$.) We also require that TRS is broken already at
the level of the classical dynamics (as it is in a quantum dot in a uniform magnetic
field). Finally, we need to relate the TRS-breaking parameter in the
stroboscopic formulation to the flux enclosed by the quantum dot. All these
issues are addressed in Secs.\ II and III before we proceed to the actual
simulation in Sec.\ IV. We conclude in Sec. V.

\section{Time-reversal-symmetry breaking in the open kicked rotator}

\subsection{Formulation of the model}

The kicked rotator is a particle 
moving along a circle, kicked periodically at time intervals 
$\tau_{0}$ \cite{Haa90,Izr90}. 
The stroboscopic time evolution of a wave function is given by the Floquet 
operator ${\cal F}$. In addition to the stroboscopic time $\tau_0$ and the
moment of inertia $I$, which we set to unity, $F$ depends on the kicking
strength $K$ and the TRS-breaking parameter $\gamma$. We require
\begin{equation}
{\cal F}(-\gamma) = {\cal F}^T(\gamma),
\label{Frec}
\end{equation}
which guarantees the reciprocity relation (\ref{Srec}) for the scattering 
matrix when we open up the model.
 
We will consider two different representations of ${\cal F}$, both of which
can be written as an $M\times M$ unitary matrix. The classical limit corresponds
to a map defined on a toroidal phase space. The difference between
the two representations is whether TRS breaking persists in the classical
limit or not. The simplest representation of ${\cal F}$ has one kick
per period. It breaks TRS quantum mechanically, but not classically. 
This would correspond to a quantum dot that encloses a flux tube, but
in which the magnetic field vanishes. A more realistic model has TRS
breaking both at the quantum mechanical and at the classical level.
We have found that we then need a minimum of three kicks per period.

\subsection{Three-kick representation}

We will mainly consider the three-kick model, so we describe it first.
In this model TRS is broken both quantum mechanically and classically.
Stroboscopic models with multiple kicks per period of different magnitude
were studied previously in the context of quantum rachets\cite{Dit00}. 
Inspired by that work, we study the time-dependent Hamiltonian
\begin{eqnarray}
H(t) &=& \frac{p^2}{2} + \frac{1}{2}V(\theta) \sum_n 
[\delta_{\epsilon} (t-n+\epsilon) + \delta_{\epsilon} (t-n-\epsilon)] \nonumber\\
&& \mbox{}+\gamma \cos(\theta) \sum_n \delta (t-n+1/3)\nonumber\\
&& \mbox{}-\gamma \cos(\theta) \sum_n \delta (t-n-1/3),
\label{Ham}
\end{eqnarray}
with $\epsilon$ an infinitesimal. The angular momentum operator
$p = -i \hbar_\text{eff} \partial_{\theta}$ is 
canonically conjugate to the angle $\theta \in [0,2\pi)$. The effective
Planck constant is $\hbar_\text{eff}=\hbar \tau_0/I$.
The potential\cite{Blu92,Tha93,Cas94,Kot03}
\begin{equation}
V(\theta) = K\cos(\pi q/2) \cos(\theta)+\frac{1}{2}K\sin(\pi q/2) \sin(2\theta)
\end{equation} 
with $q \neq 0$ breaks the parity symmetry of the model.
The form of the potential is such that in the large $K$-limit the diffusion
constant does not depend on $q$.
For $\gamma = 0$ there are two kicks per period in Eq.\ (\ref{Ham}), but since they are 
displaced by an infinitesimal amount we still call this a ``single-kick'' model.
For $\gamma \neq 0$ two more kicks appear with opposite sign at finite displacement.
We will see that this choice guarantees the reciprocity
criterion (\ref{Frec}) for the Floquet operator.

The reduction of the Floquet operator
\begin{equation}
{\cal F} = {\cal T}\exp\left[ -\frac{i}{\hbar_{\text{eff}}} \int_0^{1}H(t) dt  \right]
\end{equation}
(with ${\cal T}$ the time ordering operator)
to a discrete, finite form
is obtained only for special values of $\hbar_\text{eff}$, known
as resonances \cite{Izr90}. We have to reconsider the usual condition
for resonances in the presence of additional, TRS-breaking kicks.
Here our analysis departs from the quantum rachet analogy\cite{Dit00}.

The initial wave function $\psi(\theta)$ evolves in one period
to a final wave function $\bar{\psi}(\theta)$, given by 
\begin{eqnarray}
\bar{\psi}(\theta) & = & 
\exp(-i V(\theta)/2\hbar_\text{eff}) 
\exp(i \hbar_\text{eff} \partial_\theta^2/6) \nonumber \\
&&
\times \exp(-i \gamma\cos(\theta)/\hbar_\text{eff})
\exp(i \hbar_\text{eff} \partial_\theta^2/6) \nonumber \\
&& \times \exp(i \gamma\cos(\theta)/\hbar_\text{eff})
\exp(i \hbar_\text{eff} \partial_\theta^2/6) \nonumber \\
&&
\times \exp(-i V(\theta)/2\hbar_\text{eff}) \psi(\theta).
\label{Floq}
\end{eqnarray}
One recognizes three factors describing free propagation for $1/3$ of a period,
each followed by a kick.
The resonance condition for free propagation is 
$\hbar_\text{eff}=2\pi r/M$, with $r$ an odd integer and $M$ 
an even integer\cite{Izr90}.
The free propagation
\begin{equation}
\psi_1(\theta) = \exp(i \hbar_\text{eff} \partial_\theta^2/6) \psi(\theta)
\end{equation}
is then given by
\begin{eqnarray}
\label{Free}
\psi_1\left(\theta+{ \frac{2\pi}{3M}}n\right) =
{ \frac{1}{3M}} \sum_{m,n'=0}^{3M-1} 
\exp\left(-i \frac{2 \pi r}{3 M} m^2\right) \nonumber\\
\times \exp\left(-i m \frac{2 \pi}{3 M} (n'-n)\right)
\psi\left(\theta+{ \frac{2\pi}{3M}}n'\right). \nonumber \\ 
\end{eqnarray}
Resonance means that
the initial and final wave functions can be treated as discrete vectors
on a $3M$-point lattice, labeled by the indices $n$, $n'$. The angle $\theta$ 
is an arbitrary offset parameter. Different values of $\theta$ are not
coupled by the free propagation.
Putting together three iterations of Eq.\ (\ref{Free})
we get three independent components of $\psi(\theta+2\pi n/3M)$ 
for $n =0,1,2$ (mod $3$), each on an $M$-point lattice.

We find that the resonance property is preserved in the presence of intervening
TRS-breaking kicks, provided that $r=3$ and $M$ even, but not a multiple of $3$.
The free propagation (\ref{Free}) then is conveniently expressed in matrix notation.
The matrix acts on an $M$-component vector $\psi_{m}=\psi(\theta+2\pi m/3M)$, 
$m=0,\ldots,M-1$. We choose the arbitrary phase $\theta =0$, so that
\begin{equation}
(\psi_1)_m = \sum_{m'=0}^{M-1} (U^{\dagger} \Sigma U)_{m m'} \psi_{m'}.
\end{equation}
The matrices are defined by
\begin{eqnarray}
\Sigma_{mm'}& = & \delta_{mm'}e^{-i\pi m^{2}/M},\\
U_{mm'} & = & M^{-1/2}e^{-2\pi imm'/M}.
\end{eqnarray}
The matrix product $U^{\dagger}\Sigma U$ can be evaluated in closed form,
with the result
\begin{eqnarray}
\Pi_{mm'} &=& (U^{\dagger}\Sigma U)_{mm'} \nonumber \\
&=&M^{-1/2}e^{-i\pi/4}\exp[i(\pi/M)(m'-m)^{2}]. \nonumber \\
&& 
\label{UPUdef}
\end{eqnarray}
 
Collecting results, we find that for $\hbar_{\text{eff}}=6\pi/M$ the Floquet
operator (\ref{Floq}) is represented by an $M\times M$ unitary matrix,
of the form
\begin{subequations}
\label{Fdef}
\begin{eqnarray}
{\cal F}_{mm'} & = & (X \Pi Y^* \Pi Y \Pi X)_{mm'},\label{Fdefa}\\
Y_{mm'} & = & \delta_{mm'}e^{i(M\gamma/6\pi)\cos(2\pi m/M)},\label{Fdefb}\\
X_{mm'} & = & \delta_{mm'}e^{-i(M/12\pi)V(2\pi m/M)}.\label{Fdefc}
\end{eqnarray}
\end{subequations}
One readily verifies the reciprocity relation (\ref{Frec}).

The classical map corresponding to this quantum mechanical model is 
derived in App. A. We show there that TRS-breaking of the classical map
is broken for $\gamma\neq 0$ in the three-kick model.

\subsection{One-kick representation}

TRS breaking in the one-kick model is constructed as a formal analogy to 
the magnetic
vector potential, by adding an offset $\delta$ to the momentum
of the kicked rotator \cite{Izr86,Izr90,Blu92,Tha93,Cas94,Kot03,Shu96}. 
%This can be derived as a rotator at the resonance \cite{Wim} for 
%only discrete values of $\gamma$. For continuous $\gamma$ one has to additionaly
%enforce the periodicity of momenta.
%Here we consider another version of this model, which has both reciprocity 
%and is formulated in the angle representation. 

To obey reciprocity
\begin{equation}
{\cal F}(-\delta) = {\cal F}^T(\delta)
\label{Frec_d}
\end{equation}
for odd $M$ it is enough to symmetrize the expression of
Ref.\ \onlinecite{Izr86}
by infinitesimally splitting the kick (as it was done in Ref.\ \onlinecite{Two03}
for $\delta =0$).
For even $M$, which is more convenient for application of the fast Fourier
transform, one also needs to
redefine the lattice points in order to preserve reciprocity\cite{Gor04}.

The model takes the form
\begin{subequations}
\label{Fdef_delta}
\begin{eqnarray}
&&{\cal F}_{mm'}=(\tilde{X}\tilde{U}^{\dagger}\tilde{\Pi} \tilde{U}\tilde{X})_{mm'},\\
&&\tilde{U}_{mm'}=M^{-1/2}e^{-2\pi i(m-\frac{M-1}{2}) m'/M},\\
&&\tilde{X}_{mm'}=\delta_{mm'}e^{-i(MK/4\pi)\cos(2\pi m/M + \phi)},\\
&&\tilde{\Pi}_{mm'}=\delta_{mm'}e^{-i\pi (m-{\textstyle \frac{M-1}{2}}-\delta {\textstyle \frac{M}{2\pi}})^{2}/M}.
\end{eqnarray}
\end{subequations}
In addition to the TRS-breaking phase $\delta$ there is a phase $\phi$ to break
the parity symmetry. The reciprocity property (\ref{Frec_d}) can easily be checked.
%by making the index change $m \rightarrow -m+M-1$ in the internal sum of 
%$(\tilde{U}^{\dagger}\tilde{\Pi} \tilde{U})_{mm'}$.

The classical map corresponding to this model is also discussed in App. A. It does
not break TRS.

\subsection{Scattering matrix}

To model a pair of $N$-mode ballistic point contacts that couple the quantum dot to electron
reservoirs, we impose open boundary conditions 
in a subspace of Hilbert space represented by the indices $m_{n}^{(\alpha)}$. The
subscript $n=1,2,\ldots N$ labels the modes and  the superscript $\alpha=1,2$ labels
the leads. A $2N\times M$ projection matrix
$P$ describes the coupling to the ballistic leads. Its elements are
\begin{equation}
P_{nm}=\left\{\begin{array}{ll}
1&\text{ if}\;\;m=n\in\{m_{n}^{(\alpha)}\},\\
0&\text{ otherwise}.
\end{array}\right. \label{Wdef}
\end{equation}
The mean dwell time is $\tau_D = M/2N$ (in units of $\tau_0$).

The matrices $P$ and ${\cal F}$ together determine the scattering matrix \cite{Oss03,Jac03,Two03}
\begin{equation}
S(\varepsilon)=P[e^{-i\varepsilon}-{\cal F}(1-P^\text{T}P)]^{-1}{\cal F}P^\text{T},\label{Sdef}
\end{equation}
where $\varepsilon$ is the quasi-energy. The reciprocity condition (\ref{Frec}) of ${\cal F}$ 
implies that also $S$ satisfies the reciprocity condition (\ref{Srec}).

By grouping together the $N$ indices belonging to the same point contact, the $2N\times 2N$ matrix $S$ can 
be decomposed into 4 sub-blocks containing the $N\times N$ transmission and reflection matrices,
\begin{equation}
S=\left(\begin{array}{cc}
r&t\\t'&r'
\end{array}\right).\label{Srt}
\end{equation}
The conductance $G$ (in units of $e^2/h$, disregarding spin degeneracy)
follows from the Landauer formula
\begin{equation}
G=\text{Tr}\,tt^{\dagger}.\label{Gtt}
\end{equation}

\section{Relation with random-matrix theory}

In RMT time-reversal symmetry is broken by means of the Pandey-Mehta Hamiltonian \cite{Meh83}
\begin{equation}
H = H_0 + i \alpha A,
\end{equation}
which consists of the sum of a real symmetric matrix $H_0$ and a real antisymmetric matrix $A$
with imaginary weight $i \alpha$. We denote by $M_H$ the dimensionality of the Hamiltonian 
matrix. The two matrices $H_0$ and $A$ are independently distributed with the same Gaussian
distribution. The variance $\nu^2=\left<(H_0)_{ij}^2\right>=\left<A_{ij}^2\right>$ $(i \neq j)$ determines the mean 
level spacing $\Delta = \pi \nu/\sqrt{M_H}$ at the center of the spectrum for $M_H\gg 1$ and 
$\alpha \ll 1$.

To lowest order in perturbation theory the energy levels $E_i(\alpha)$ depend on
the TRS-breaking parameter $\alpha$ according to
\begin{equation}
\delta E_i = \alpha^2 \sum_{j\neq i} \frac{A_{ij}^2}{E_i-E_j},
\label{Ei}
\end{equation}
with $ \delta E_i = E_i(\alpha)-E_i(0)$ and $E_i \equiv E_i(0)$. The characteristic
value $\alpha_c$ is determined by the mean absolute curvature,
\begin{equation}
\alpha_c \equiv  \left( \frac{1}{\Delta} 
\left<
\left| \frac{d^2 E_i}{d \alpha^2} \right|_{\alpha=0}
\right>
\right)^{-1/2}.
\label{alphac}
\end{equation}
From Eq.\ (\ref{Ei}) we deduce that $\alpha_c \simeq \Delta/\nu \simeq 1/\sqrt{M_H}$,
up to a numerical coefficient of order unity. A numerical calculation gives
\begin{equation}
\alpha_c \sqrt{M_H} \equiv \kappa_{\text{RMT}} = 1.27.
\end{equation}

A real magnetic field $B$ is related to the parameter $\alpha$ of RMT by 
\begin{equation}
\label{b/bc}
B/B_c = \alpha/ \alpha_c,
\end{equation}
where $B_c$ is determined by the level curvature according to Eq.\ (\ref{bc}). 
For a ballistic two-dimensional billiard (area $A$, Fermi velocity $v_{\text{F}}$) 
with a chaotic classical  dynamics, one has \cite{Bee97,Boh95}
\begin{equation}
B_c = c \frac{h}{e A} ( \Delta \sqrt{A}/\hbar v_F)^{1/2},
\end{equation}
with $c$ a numerical coefficient that depends only on the shape of the 
billiard. The field $B_c$ corresponds to a flux through the quantum dot
of order $(h/e)\sqrt{\tau_{\text{erg}}\Delta/\hbar} \ll h/e$, with the 
ergodic time $\tau_{\text{erg}}$ being the time it takes an electron to explore 
the available phase space in the quantum dot.

The analogue of Eqs.\ (\ref{bc}) and (\ref{b/bc})  for the quantum kicked rotator 
considered here is 
\begin{equation}
\label{gamma/gammac}
\gamma/\gamma_c = \alpha/ \alpha_c, \; \gamma_c \equiv 
\left( \frac{M}{2\pi} 
\left< 
\left| \frac{d^2 \phi_i}{d \gamma^2} \right|_{\gamma=0}
\right>
\right)^{-1/2}.
\end{equation}
Here $\gamma$ is the TRS-breaking parameter in the three-kick model. The same relation applies 
to the one-kick model, with $\gamma$, $\gamma_c$ replaced by $\delta$, $\delta_c$.

\begin{figure}
\includegraphics[width=7cm]{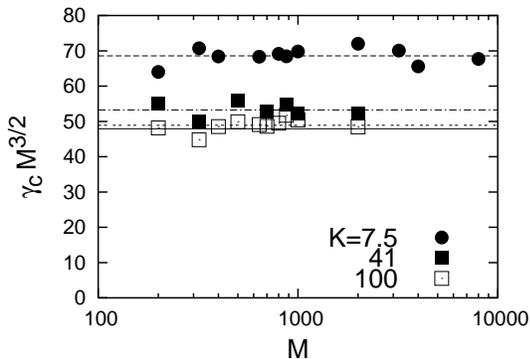}
\caption{
The critical value $\gamma_c$ of the TRS-breaking parameter in the closed three-kick model is
presented for different system sizes at fixed $K$. The parity-breaking parameter is $q=0.2$. 
The solid line shows the large-$K$ limit 
(\ref{Kinf_gm}). The dashed lines are averages over $M$ of the numerical data.
\label{rmt_fig1a}
}
\end{figure}

\begin{figure}
\includegraphics[width=7cm]{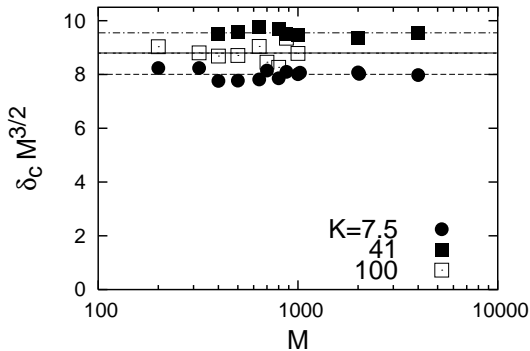}
\caption{
Same as Fig.\ \ref{rmt_fig1a}, but now for the closed one-kick model.
The parity-breaking parameter is
$\phi=0.2\, \pi$.  The solid line shows the large-$K$ limit (\ref{Kinf_dl}).
\label{rmt_fig1b}
}
\end{figure}

To complete the correspondence between the kicked rotator, RMT, and the real quantum
dot, we need to determine the two characteristic values $\gamma_c$ and $\delta_c$.
In App. B we present an analytical calculation deep in the chaotic regime
($ K \rightarrow \infty $), according to which
\begin{eqnarray}
\lim_{K \rightarrow \infty} \gamma_c = 12 \pi M^{-3/2} \kappa_{\text{RMT}} = 47.9\; M^{-3/2}, \label{Kinf_gm}\\
\lim_{K \rightarrow \infty} \delta_c = 4 \sqrt{3} M^{-3/2} \kappa_{\text{RMT}} = 8.80\; M^{-3/2}. \label{Kinf_dl}
\end{eqnarray}
In Figs.\ \ref{rmt_fig1a} and \ref{rmt_fig1b} we show a numerical calculation for finite $K$, which confirms these
analytical large-$K$ limits.

In the open system the characteristic field scale for TRS-breaking is increased by a factor
$\sqrt{g}$, with $g$ the conductance of the point contacts. We consider ballistic $N$-mode
point contacts, so that $g=N$, measured in units of $e^2/h$. The conductance $G(B)$ of the 
quantum dot is also measured in units of $e^2/h$. According
to RMT, the weak localization magnetoconductance is given by \cite{Plu94,Fra95}
\begin{equation}
G(B) = \frac{N}{2} - \frac{1}{4} \left[1+(2\kappa_{\text{RMT}} N^{-1/2} B/B_c)^2\right]^{-1}.
\end{equation}
For the quantum kicked rotator we would therefore expect a weak localization peak
in the conductance given by
\begin{equation}
G(\gamma) = G_\infty - \frac{1}{4} \left[1+(2\kappa_{\text{RMT}} N^{-1/2} \gamma/\gamma_c)^2\right]^{-1},
\label{WLpeak}
\end{equation}
in the three-kick model. We define the weak localization correction
$\delta G(\gamma) = G(\gamma) - G_\infty$, 
with $G_{\infty}$ the conductance at fully broken TRS. The expression in the
one-kick model is similar, with $\gamma/\gamma_c$ replaced by $\delta/\delta_c$.

In the large-$K$ limit we can use the analytical expressions (\ref{Kinf_gm}) and (\ref{Kinf_dl})
for $\gamma_c$ and $\delta_c$ to obtain
\begin{eqnarray}
\lim_{K \rightarrow \infty} \delta G(\gamma) = - \textstyle{\frac{1}{4}} [1+(M^{3/2} N^{-1/2} \gamma/6 \pi)^2]^{-1}, \\
\lim_{K \rightarrow \infty} \delta G(\delta) = - \textstyle{\frac{1}{4}} [1+(M^{3/2} N^{-1/2} \delta/2 \sqrt{3})^2]^{-1}.
\end{eqnarray}
In App. C we show how these two results are consistent with a semiclassical calculation.

\section{Numerical results \label{sec3}}

The numerical technique we use to calculate the conductance was described in Refs.\ \onlinecite{Two03}
and \onlinecite{Two04}. The calculation of the scattering matrix (\ref{Sdef}) is performed
efficiently by use of an iterative procedure and the fast-Fourier-transform algorithm.
We need to average over many
system realizations (varying lead positions and quasi-energies) to suppress
statistical fluctuations. In addition, we need several points
to plot the $\gamma$-dependence. This makes the calculation for
large $M$ more time consuming than earlier studies
of universal conductance fluctuations in the same model at zero magnetic field\cite{Two04,Jac04}.

First we present in Figs. 3 and 4 results for the weak localization correction $\delta G$
in the three-kick model
as a function of the TRS-breaking parameter $\gamma$.
The data are obtained by averaging over 
$40$ lead positions and $80$ quasi-energies. 
The parameter $\gamma_c$ was calculated for the closed model using Eq.\ (\ref{gamma/gammac}), and the
resulting RMT prediction (\ref{WLpeak}) is also shown (dotted curve).
%Each data set for every system size $M$ was separately fitted with 
%the formula (\ref{WLpeak})  to obtain $G_\infty$. 
%We see a good agreement with the RMT prediction combined with $\gamma_c$ calculated
%for the closed model according to Eq.\ (\ref{gamma/gammac}).

\begin{figure}
\includegraphics[width=7cm]{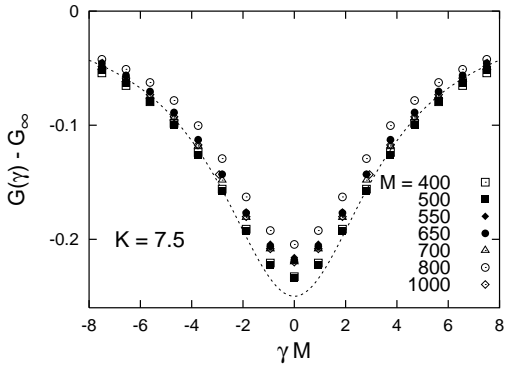}
\caption{
Dependence of the average conductance on the TRS-breaking parameter $\gamma$.
The three-kick model is characterized by $K=7.5$, $q=0.2$, and $\tau_{D}=M/2N=25$. 
The dotted line shows the RMT prediction (\ref{WLpeak}), with $\gamma_c$ 
calculated from the mean level curvatures (Fig.\ \ref{rmt_fig1a}).
\label{fig1}
}
\end{figure}

\begin{figure}
\includegraphics[width=7cm]{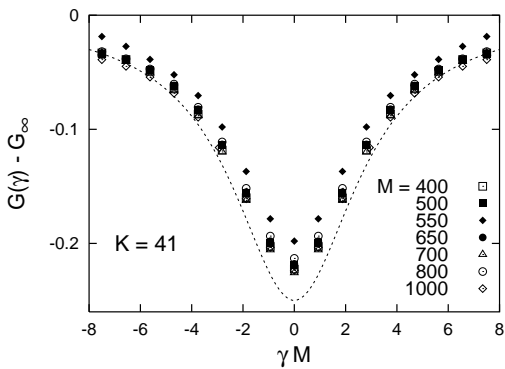}
\caption{
Same as Fig.\ \ref{fig1}, but for $K=41$.
\label{fig2}
}
\end{figure}

\begin{figure}
\includegraphics[width=7cm]{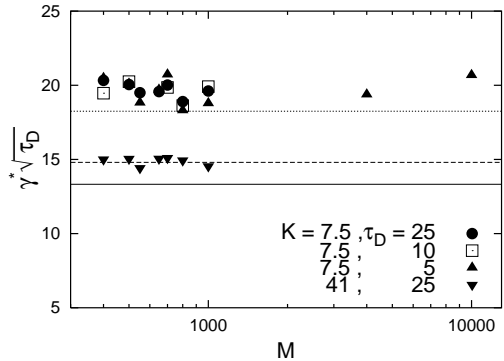}
\caption{
Dependence of the crossover parameter $\gamma^*$ on the system size. 
The data are 
obtained by fitting the Lorentzian (\ref{fitG}) to the numerical data
of Figs.\ \ref{fig1} and \ref{fig2}.
The solid line shows the large $K$-limit (\ref{Kinf_gm}). The dotted
lines are the RMT prediction for $K=7.5$ and $K=41$, using $\gamma_c$ 
found from the level curvatures in the closed model (Fig.\ \ref{rmt_fig1a}).
\label{fig3}
}
\end{figure}

To compare the simulation with RMT in more detail we have fitted a Lorentzian
\begin{equation}
\delta G = - \textstyle{\frac{1}{4}} [1+(M \gamma/\gamma^*)^2]^{-1}
\label{fitG}
\end{equation}
to each data set. This is the RMT result (\ref{WLpeak}) if 
$\gamma^*=\gamma^*_{\rm RMT} \equiv \gamma_c M^{3/2}/(2\sqrt{2\tau_D}\kappa_{\text{RMT}})$.
The large $K$-limit is 
\begin{equation}
\lim_{K\rightarrow\infty} \gamma^*_{\rm RMT}=6\pi/\sqrt{2\tau_D}.
\label{gm_star}
\end{equation}
In Fig.\ \ref{fig3}  we plot the fitted crossover parameter $\gamma^*$
as a function of $M$ for fixed dwell time. The plot confirms the scaling 
with $\tau_D^{-1/2} \propto g^{-1/2}$, and also shows good agreement with the 
values of $\gamma^*_{\rm RMT}$ calculated from the mean level curvature (dotted lines).

\begin{figure}
\includegraphics[width=7cm]{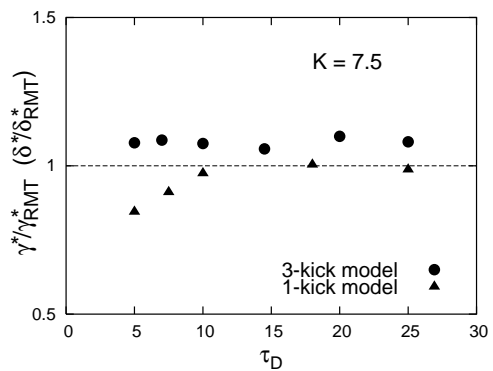}
\caption{
Dependence of the ratio $\gamma^*/\gamma_{\rm RMT}^*$ for the three-kick model
and the ratio $\delta^*/\delta_{\rm RMT}^*$ for the one-kick model
on the dwell time $\tau_D$. Data points for a given dwell time are obtained
by averaging over system sizes in the range from $200$ to $1000$.
}
\label{fig4}
\end{figure} 

We also performed numerical calculations for the one-kick model.
The crossover scale $\delta^*$ extracted from a Lorentzian fit to the 
weak-localization peak was compared with the value 
$\delta^*_{\rm RMT}=\delta_c M^{3/2}/(2\sqrt{2\tau_D}\kappa_{\rm RMT})$ predicted 
by the mean level curvature. 
The large $K$-limit of this value is 
\begin{equation}
\lim_{K\rightarrow\infty} \delta^*_{\rm RMT}= \sqrt{6}/\sqrt{\tau_D}.
\label{dl_star}
\end{equation}
We show in Fig.\ \ref{fig4} 
the ratio $\delta^*/\delta_{\rm RMT}^*$
for the one-kick model, as well as the ratio $\gamma^*/\gamma_{\rm RMT}^*$
for the three-kick model. 
The ratio is close to unity for both models if the dwell time is sufficiently
large. At the smallest $\tau_D$ there is some deviation from unity in the
one-kick model.

\begin{figure}
\includegraphics[width=7cm]{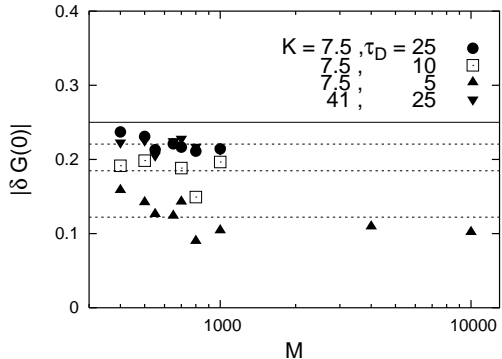}
\caption{
Dependence of $\delta G(0)$ 
on the system size $M$ for several dwell times. Dashed lines show averages
over system size.
}
\label{fig5}
\end{figure} 

\begin{figure}
\includegraphics[width=7cm]{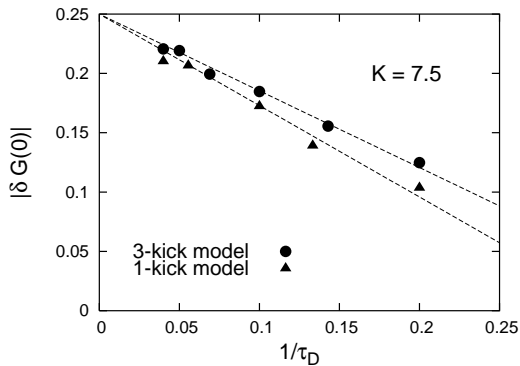}
\caption{
Dependence of the amplitude of the weak localization peak $\delta G(0)$ 
(averaged over several system sizes) on the dwell time $\tau_D$. Dashed lines
show a linear dependence on $1/\tau_D$, extrapolated to the RMT value
$|\delta G(0)|=1/4$.
}
\label{fig6}
\end{figure} 

The magnitude of the weak localization peak in Figs.\ \ref{fig1} and \ref{fig2}
shows a small (about 10\%) discrepancy with the RMT prediction. We attribute
this to non-ergodic, short-time trajectories. We show in Fig.\ \ref{fig6}
the dependence of the magnitude of the weak localization peak $\delta G(0)$
on the dwell time. The results
suggest that $\delta G(0)+\frac{1}{4}\propto 1/\tau_D$, a deviation 
from RMT to be expected from the Thouless energy scale (which is $ \propto 1/\tau_D$). 
The deviation from unity in Fig.\ \ref{fig4} has presumably the same origin.

We could determine the $M$-dependence of 
$\gamma^*$ and $\delta G(0)$ up to
$M=10^4$ (for $K=7.5$ and $\tau_D=5$). 
The motivation for extending the calculation to large system sizes is to search for effects 
of the Ehrenfest time\cite{Ale96,Ada03}. 
Although the Ehrenfest time $\tau_E \approx 3.8$ (estimating as in 
Ref.\ \onlinecite{Two03}) was comparable to $\tau_D=5$, we did not find any 
systematic $M$-dependence in $\gamma^*$ or $\delta G(0)$, cf. Figs.\ \ref{fig3} and \ref{fig5}.

\section{Conclusions}

In conclusion, we have studied time-reversal symmetry breaking in quantum chaos
through its effect on weak localization. We have found an overall good agreement
between the universal predictions of random-matrix theory and the results for a
specific quantum mechanical model of a chaotic quantum dot. In particular, the
scaling $\propto g^{-1/2}$ of the crossover magnetic field with the point 
contact conductance $g$ is confirmed over a broad parameter range.

Deviations from RMT that we have observed scale inversely proportional with the mean dwell
time $\tau_D$, consistent with an explanation in terms of non-ergodic short-time 
trajectories. These deviations therefore have a classical origin.

More interesting deviations of a quantum mechanical origin have been predicted \cite{Ale96,Ada03}
in relation with the finite Ehrenfest time $\tau_E$. This is the time scale on which a
wave packet of minimal initial dimension spreads to cover the entire 
available phase space. The theoretical prediction is that the weak localization peak 
$\delta G(0) \propto e^{-\tau_E/\tau_D}$ should decay exponentially once $\tau_E$
exceeds $\tau_D$. Our simulation extends up to $\tau_E \simeq \tau_D$, but shows
no sign of this predicted decay. This is consistent with the explanation advanced
by Jacquod and Sukhorukov \cite{Jac04} for the insensitivity of universal conductance 
fluctuations to a finite Ehrenfest time. As pointed out in Ref. \onlinecite{Two04}
the same explanation also implies that weak localization should not depend on the
relative magnitude of $\tau_E$ and $\tau_D$.

Because our simulation could not be extended to the regime $\tau_E > \tau_D$, this
final conclusion remains tentative. It might be that the exponential suppression
of $\delta G(0)$ does exist, but that our system was simply too small to see it.

\acknowledgments

We benefitted from discussions with M. C.\ Goorden, Ph.\ Jacquod, and H.\ Schomerus.

This work was supported by the Dutch Science Foundation NWO/FOM. J.T. 
acknowledges the financial support provided through the European 
Community's Human Potential  Programme under contract HPRN--CT--2000-00144, 
Nanoscale Dynamics.

\appendix

\section{Classical map}

Here we derive the classical map that is associated with the quantum mechanical
Floquet operator of the kicked rotator with broken TRS. We consider separately
the three-kick and one-kick representation.

\subsection{Three-kick representation}

We seek the classical limit of the Floquet operator (\ref{Fdef}). We consider
the classical motion from $\theta_0$ at $t=0$ to $\theta_T$ at $t=T$ 
(in units of $\tau_0$). Intermediate values of the coordinate are denoted by $\theta_t$,
$t=0,1,...,T$. The classical action ${\cal S}$ is the sum
\begin{equation}
{\cal S} = \sum_{t=0}^{T-1} S(\theta_{t+1},\theta_t).
\end{equation}
Following the general method of Ref.\ \onlinecite{Shu97} we derive
\begin{equation}
S(\theta',\theta) = S_c(\theta',\theta_2) + S_b(\theta_2,\theta_1) + S_a(\theta_1,\theta),
\end{equation}
\begin{eqnarray}
S_a(\theta_1,\theta) & = & {\textstyle \frac{3}{2}} (\theta_1-\theta+2\pi\sigma_{\theta_1})^2 - 6\pi\sigma_{p_1} \theta_1 \nonumber \\
&& \mbox{} + \gamma \cos(\theta_1)- {\textstyle \frac{1}{2}} V(\theta),\\
S_b(\theta_2,\theta_1) & = & {\textstyle \frac{3}{2}} (\theta_2-\theta_1+2\pi\sigma_{\theta_2})^2 - 6\pi\sigma_{p_2} \theta_2, \\
S_c(\theta',\theta_2) & = & {\textstyle \frac{3}{2}} (\theta'-\theta_2+2\pi\sigma_{\theta'})^2 - 6\pi\sigma_{p'} \theta' \nonumber \\
&& \mbox{} - \gamma \cos(\theta_2) - {\textstyle \frac{1}{2}} V(\theta).
\end{eqnarray}
The integers $\sigma_\theta,\sigma_p$ are the winding numbers of a classical 
trajectory on a torus with $\theta\in [0,2\pi)$ and $p\in[0,6\pi)$.
The map equations are derived from
\begin{eqnarray}
p_1 &=&\frac{\partial }{\partial \theta_1} S_a(\theta_1,\theta),\quad
p =-\frac{\partial}{\partial \theta} S_a(\theta_1,\theta), \label{pmapa} \\
p_2 &=&\frac{\partial }{\partial \theta_2} S_b(\theta_2,\theta_1),\quad
p_1 =-\frac{\partial}{\partial \theta_1} S_b(\theta_2,\theta_1),\label{pmapb} \\
p' &=&\frac{\partial }{\partial \theta'} S_c(\theta',\theta_2),\quad
p_2 =-\frac{\partial}{\partial \theta_2} S_c(\theta',\theta_2). \label{pmapc}
\end{eqnarray}
Eqs.\ (\ref{pmapa}-\ref{pmapc}) are equivalent to the following set of 6 equations
that map initial coordinates $(\theta,p)$ onto final coordinates $(\theta',p')$
after one period:
\begin{eqnarray}
&& \left\{
\begin{array}{l}
\theta_1 = \theta +  p/3 - V'(\theta)/6 - 2 \pi\sigma_{\theta_1},\\
p_1 = p - \gamma \sin\theta_1 - V'(\theta)/2 - 6 \pi\sigma_{p_1},
\end{array}
\right. \\
&& \left\{
\begin{array}{l}
\theta_2 = \theta_1 +  p_1/3 - 2 \pi\sigma_{\theta_2}, \\
p_2 = p_1 - 6 \pi\sigma_{p_2},
\end{array}
\right. \\
&& \left\{
\begin{array}{l}
\theta' = \theta_2 +  p_2/3 + \gamma \sin \theta_2 /3 - 2 \pi\sigma_{\theta'},\\
p' = p_2 + \gamma \sin \theta_2 - V'(\theta')/2 - 6 \pi\sigma_{p'}.
\end{array}
\right.
\end{eqnarray}
We denote $V'=dV/d\theta$.
Winding numbers of a trajectory on the torus in phase space $(\theta,p)$ are 
denoted by $\sigma_\theta$, $\sigma_p$.
These integers are determined
by the requirement  that $\theta,\theta_1,\theta_2,\theta'\in [0,2\pi)$ and 
$p,p_1,p_2,p'\in [0,6\pi)$. TRS for a classical map means that
the point $(\theta',-p')$ maps to $(\theta,-p)$.
This property is satisfied for $\gamma=0$, but not for $\gamma \neq 0$.
TRS is broken at the classical level in the three-kick model.

\subsection{One-kick representation}

We now seek the classical limit of the Floquet operator (\ref{Fdef_delta}).
The classical action $S$ after one kick is
\begin{eqnarray}
S(\theta',\theta) & = & {\textstyle \frac{1}{2}} (\theta'-\theta+2\pi\sigma_\theta)^2 - 2\pi\sigma_p \theta' \nonumber \\
&& \mbox{} + \delta (\theta'-\theta+2\pi\sigma_\theta) \nonumber \\
&& \mbox{} - {\textstyle \frac{1}{2}} K [\cos(\theta+\phi)+\cos(\theta'+\phi)].
\end{eqnarray}
%The integer parameters $\sigma_\theta,\sigma_p$ are the winding numbers of a classical 
%trajectory correspondingly in $\theta\in[0,2\pi[$ and $p\in[0,2\pi[$ direction.
The map equations are derived from
\begin{equation}
p'=\frac{\partial }{\partial \theta'} S(\theta',\theta),\quad
p =-\frac{\partial}{\partial \theta}S(\theta',\theta).
\end{equation}
The mapping of initial coordinates $(\theta,p)$ onto final ones $(\theta',p')$
after one kick is then
\begin{equation}
\label{map_mar}
\left\{
\begin{array}{rcl}
\theta' &=& \theta + p + {\textstyle \frac{1}{2}}K \sin(\theta+\phi) - \delta - 2\pi\sigma_\theta,\\
p' &=& p + {\textstyle \frac{1}{2}} K [\sin(\theta+\phi) + \sin(\theta'+\phi)] - 2\pi\sigma_p.
\end{array}
\right.
\end{equation}
%Both $\sigma_\theta$ and $\sigma_p$ are determined uniquely by the requirement 
%that $\theta'\in [0,2\pi[$ and $p'\in [0,2\pi[$. 

The canonical transformation
$p-\delta \rightarrow \tilde{p}$, $\theta+\phi \rightarrow \tilde{\theta}$
brings the map to an equivalent form
\begin{equation}
\label{map_inv}
\left\{
\begin{array}{rcl}
\tilde{\theta}' &=& \tilde{\theta} + \tilde{p} + {\textstyle \frac{1}{2}}K \sin\tilde{\theta}  - 2\pi\sigma_\theta,\\
\tilde{p}' &=& \tilde{p} + {\textstyle \frac{1}{2}} K (\sin\tilde{\theta} + \sin\tilde{\theta}') - 2\pi\sigma_p.
\end{array}
\right.
\end{equation}
This form is manifestly invariant under the transformation that maps
$(\tilde{\theta}',-\tilde{p}')$ onto $(\tilde{\theta},-\tilde{p})$ for any value of
$\phi$ and $\delta$. Hence TRS is not broken at the classical level in the
one-kick model.

%This symmetry is not preserved in a quantum model defined on a tori, becouse the
%cannonical offset of momentum by arbitrary amount $\delta$ does not 
%coincide with descreete momentum eigenvalues.

\section{Derivation of Eqs.\ (\ref{Kinf_gm}) and (\ref{Kinf_dl})}

In the large-$K$ limit the level curvature in the kicked rotator can be 
related to the level curvature in the Pandey-Mehta Hamiltonian. This leads to the
relations (\ref{Kinf_gm}) and (\ref{Kinf_dl}) between the TRS breaking parameters
$\gamma$ (three-kick model) and $\delta$ (one-kick model),
on the one hand, and the Pandey-Mehta parameter $\alpha$, on the other hand.

Perturbation theory for eigenphases $\phi_i(\delta \gamma)$ of a unitary matrix 
${\cal F}(\delta \gamma)$ gives the series expansion
\begin{eqnarray}
\phi_i(\delta \gamma) &=& \phi_i + W_{ii} \delta \gamma + 
\textstyle{\frac{1}{2}} \sum_{j \neq i} |W_{ij}|^2 (\delta \gamma)^2 \text{cotan} \frac{\phi_i-\phi_j}{2}  \nonumber\\
&&\mbox{} +\textstyle{\frac{1}{2}} V_{ii} (\delta \gamma)^2.
\label{Per}
\end{eqnarray}
Here $\phi_i$ denotes an eigenphase of 
${\cal F}(0)=U \text{diag}(e^{i\phi_1},\ldots,e^{i\phi_M}) U^{\dagger}$.
The Hermitian matrices $W$ and $V$ are defined by 
$W= U(- i {\cal F}^{\dagger} \partial_\gamma {\cal F}|_{\gamma=0})U^{\dagger}$,
$V = \partial_\gamma W|_{\gamma=0}$. Due to reciprocity of ${\cal F}$ we find $W_{ii}=0$.
%${\cal F}(\gamma + \delta \gamma)={\cal F}(\gamma) e^{i\delta \gamma W + i(\delta \gamma)^2 V}$
%for infinitesimal increment $\delta \gamma$ (unitarity of ${\cal F}$ is preserved). 
%Expandindg to second order in $\gamma$ we find
%$W=-i {\cal F}^{\dagger} \partial_\gamma {\cal F}$ and $V=\partial_\gamma W$,  
For the three-kick model (\ref{Fdef}) the operators $W$, $V$ are
\begin{equation}
W = \frac{M}{6 \pi} U X^{\dagger} \Pi^{\dagger} Y^{\dagger} \Pi^{\dagger}
(-C \Pi + \Pi C)  Y \Pi X U^{\dagger},
\end{equation}
\begin{equation}
V = i (\frac{M}{6 \pi})^2  U X^{\dagger} \Pi^{\dagger} Y^{\dagger}
(C \Pi^{\dagger} C \Pi - \Pi^{\dagger} C \Pi C) Y \Pi X U^{\dagger},
\end{equation}
where $C_{mm'}=\delta_{mm'} \cos (2\pi m/M)$. We assume that for strongly chaotic
systems ($K \gg 1$) the matrix elements $W_{ij}$ and $V_{ii}$ 
are random Gaussian numbers independent of the eigenphases. Average diagonal elements calculated in the 
three kick  model at $\gamma =0$ are
$\left< V_{ii} \right> = \text{Tr} V/M=0$ and $\left< W_{ii} \right> = \text{Tr} W/M=0$. 
The variance of the off-diagonal elements is $\left< |W_{ij}|^2 \right> = \text{Tr} WW^{\dagger}/M^2=M/(6\pi)^2$.

For the one-kick model (\ref{Fdef_delta}) the operators $W$, $V$ are
\begin{equation}
W =  U X^{\dagger} \tilde{U}^{\dagger} D \tilde{U} X U^{\dagger},\quad
V = - \frac{1}{2\pi} M,
\end{equation}
with $D_{mm'}=\delta_{mm'} ( m + 1/2 - M/2 - \delta M/2 \pi )$. Average diagonal elements  
at $\delta = 0$ are
$\left< V_{ii} \right> = \text{Tr} V/M=-M/2\pi$ and $\left< W_{ii} \right> = \text{Tr} W/M=0$. 
The variance of the off-diagonal elements is $\left< |W_{ij}|^2 \right> = \text{Tr} WW^{\dagger}/M^2=M/12$.

For $K\gg 1$ the eigenphases $\phi_i$ are distributed randomly in the circular ensemble, which
is locally equivalent to the Gaussian ensemble\cite{Haa90}.
We expand Eq.\ (\ref{Per}) for small eigenphases
difference, compare with Eq. (\ref{Ei}) and substitute the variances of matrix elements calculated
above. For the one-kick model we drop terms with $V_{ii}$ as they are of order $1/M$.
We finally arrive at Eqs.\ (\ref{Kinf_gm}) and (\ref{Kinf_dl}).
%\begin{eqnarray}
%\lim_{K \rightarrow \infty} \gamma_c = 12 \pi M^{-3/2} \kappa_{\text{RMT}}, \\
%\lim_{K \rightarrow \infty} \delta_c = 4 \sqrt{3} M^{-3/2} \kappa_{\text{RMT}}.
%\end{eqnarray}
 
The explicit formula for the Pandey-Mehta parameter $\alpha$ describing the kicked
rotator at large $K$ is
\begin{equation}
\label{al3}
\alpha \sqrt{M_{\text{H}}} = \frac{\gamma M^{3/2}}{12 \pi}
\end{equation}
for the three-kick model. The corresponding formula for the one-kick model is
\begin{equation}
\label{al1}
\alpha \sqrt{M_{\text{H}}} = \frac{\delta M^{3/2}}{4 \sqrt{3}}.
\end{equation}

\section{Semiclassical derivation of the weak localization peak}

We present a semiclassical derivation of the weak localization peak,
adopting the method of Ref.\ \onlinecite{Bar93} to the case of the kicked
rotator. 
The method can not be used to determine
the amplitude $\delta G(0)$,
but we use it for the crossover scale.
This serves as an independent check for the scaling
predicted by RMT.

The action difference in the three-kick model for a pair of trajectories related by TRS 
is calculated as follows. The action
${\cal S}_0$ for a trajectory with initial coordinate $\theta_0$ and final coordinate $\theta_T$ 
at $\gamma=0$ is compared
with the action ${\cal S}$ for a 
trajectory with the same initial and final coordinates, but
at small $\gamma$. The result of linear expansion in $\gamma$ is
\begin{equation}
\Delta S = {\cal S} - {\cal S}_0 = \gamma \sum_{t} [\cos\theta_1(t)-\cos\theta_2(t)],
\end{equation}
where periods are numbered by $t=0,1,\ldots,T-1$ and $\theta_1(t), \theta_2(t)$ 
denote the coordinate of the particle when TRS-breaking kicks are applied.

The weak localization correction is 
\begin{equation}
\delta G \propto \left< \exp (2i \Delta S/\hbar_{\text{eff}})\right>, 
\end{equation}
where the average is taken with respect to all trajectories connecting
initial to final coordinates. Approximating the distribution of the phase 
difference $\Delta S$ for a single step by a Gaussian,
and taking the continuum limit of exponential dwell-time probability
$P(t) \propto e^{-t/\tau_D}$, we derive
\begin{equation}
 \delta G \propto [1+(M\gamma/\gamma^*)^2]^{-1},\quad (\gamma^*)^2 = 2 \hbar_{\rm eff}^2/(\tau_D \nu), 
\end{equation}
with $\nu$ being the variance of $\Delta S/\gamma$ for a single step.
The result $\nu=1$ for large $K$ (and large $\tau_D$) is 
obtained by averaging over random initial points in the whole phase space.
We thus find Eq. (\ref{gm_star}), the same result as the one obtained in RMT.
%\begin{equation}
%\label{gammac}
%\gamma^*=6\pi/\sqrt{2 t_D}.
%\end{equation}

The action difference for a pair of symmetry related trajectories
in the one-kick model is 
\begin{equation}
\Delta S = {\cal S} - {\cal S}_0 = \delta \sum_{t}[\theta'(t)-\theta(t)+2\pi \sigma_\theta(t)],
\label{DS_delta}
\end{equation}
to linear order in $\delta$. This leads to
\begin{equation}
 \delta G \propto [1+(M\delta/\delta^*)^2]^{-1},\quad (\delta^*)^2 = 2 \hbar_{\rm eff}^2/(\tau_D \nu).
\end{equation}
By averaging over random initial points in the whole phase space for large $K$ and $\tau_D$
we find $\nu = 4\pi^2/3$. Hence we obtain Eq.\ (\ref{dl_star}), the result of RMT.
%We get
%\begin{equation}
%\label{deltac}
%\delta^*=\sqrt{6}/\sqrt{t_D}.
%\end{equation}


\begin{thebibliography}{99}

\bibitem{Haa90} F. Haake, {\em Quantum Signatures of Chaos} (Springer, Berlin, 1992).

\bibitem{Bee97} C. W. J. Beenakker, Rev.\ Mod.\ Phys.\ {\bf 69}, 731 (1997).

\bibitem{Guh98} T. Guhr, A. M\"uller-Groeling, and H. A. Weidenm\"uller, Phys. Rep.
{\bf 299}, 190 (1998).
 
\bibitem{Alh00} Y. Alhassid, Rev. Mod. Phys. {\bf 72}, 895 (2000).

\bibitem{Boh95} O. Bohigas, M. G. Giannoni, A. M. Ozorio de Almeida, and C. Schmit,
Nonlinearity {\bf 8}, 203 (1995).

\bibitem{Yan95} Z. D. Yan and R. Harris, Europhys. Lett. {\bf 32}, 437 (1995).

\bibitem{Shu97} P. Shukla and A. Pandey, Nonlinearity {\bf 10}, 979 (1997).

\bibitem{Bar93} H. U. Baranger, R. A. Jalabert, and A. D. Stone, Phys. Rev. Lett. {\bf 70},
3876 (1993); Chaos {\bf 3}, 665 (1993).

\bibitem{Plu94} Z. Pluhar, H. A. Weidenm\"uller, J. A. Zuk, and C. H. Lewenkopf, 
Phys. Rev. Lett. {\bf 73}, 2115 (1994).

\bibitem{Efe95} K. B. Efetov, Phys. Rev. Lett. {\bf 74}, 2299 (1995).

\bibitem{Fra95} K. Frahm, Europhys. Lett. {\bf 30}, 457 (1995); 
K. Frahm and J.-L. Pichard, J. Phys. I {\bf 5}, 847 (1995).

\bibitem{Oss03} A. Ossipov, T. Kottos, and T. Geisel, Europhys.\ Lett.\ {\bf 62}, 719 (2003).

\bibitem{Jac03} Ph. Jacquod, H. Schomerus, and C.W.J. Beenakker,
Phys.\ Rev.\ Lett.\ {\bf 90}, 207004 (2003).

\bibitem{Two03} J. Tworzyd\l o, A. Tajic, H. Schomerus, and C. W. J. Beenakker,
Phys.\ Rev.\ B {\bf 68}, 115313 (2003).

\bibitem{Izr86} F. M. Izrailev, Phys. Rev. Lett. {\bf 56}, 541 (1986).

\bibitem{Izr90} F. M. Izrailev, Phys.\ Rep.\ {\bf 196}, 299 (1990).

\bibitem{Blu92} R. Bl\"umel and U. Smilansky, Phys.\ Rev.\ Lett.\ {\bf 69}, 217 (1992).

\bibitem{Tha93} M. Thaha, R. Bl\"umel, and U. Smilansky, Phys.\ Rev.\ E\ {\bf 48}, 1764 (1993).

\bibitem{Car98} T. O. de Carvalho, J. P. Keating, and J. M. Robbins, J. Phys. A {\bf 31}, 5631 (1998).%cat map with borken symm.

\bibitem{Dit00} T. Dittrich, R. Ketzmeric, M.-F. Otto, and H. Schanz, Ann. Phys. (Leipzig) {\bf 9}, 755 (2000).%rachets, periodicity
%in p enforced --> condition for h like at resonance

\bibitem{Jon03} P. H. Jones, M. Goonasekera, H. E. Saunders-Singer, and D. R. Meacher, quant-ph/0309149. %kicked rot, diffusion assym.

\bibitem{Jonck03} T. Jonckheere, M. R. Isherwood, and T. S. Monteiro, Phys. Rev. Lett. {\bf 91}, 253003 (2003).  %kicked rot, experiment

%\bibitem{Sch01} H. Schanz, M.-F. Otto, R. Ketzmeric, and T. Dittrich, nlin.CD/0011038. %irrelevant: uses asym. kinet. energy

\bibitem{Cas94} G. Casati, R. Graham, I. Guarneri, and F. M. Izrailev, Phys.\ Lett.\ A\ {\bf 190}, 159 (1994).

\bibitem{Kot03} T. Kottos, A. Ossipov, and T. Geisel, Phys.\ Rev.\ E {\bf 68}, 066215 (2003).

\bibitem{Shu96} P. Shukla, Phys.\ Rev.\ E\ {\bf 53}, 1362 (1996).

%\bibitem{Wim03} S. Wimberger, Nonlinearity {\bf 16}, 1381 (2003).

%\bibitem{Hen03} H. Schomerous (private communication).

\bibitem{Gor04} M. C. Goorden and Ph. Jacquod (private communication).

\bibitem{Meh83} M. L. Mehta and A. Pandey, J. Phys. A {\bf 16}, 2655 (1983).

\bibitem{Two04} J. Tworzyd\l o, A. Tajic, and C. W. J. Beenakker, Phys. Rev. B {\bf 69}, 165318 (2004).

\bibitem{Jac04} Ph. Jacquod and E. V. Sukhorukov, Phys. Rev. Lett. {\bf 92}, 116801 (2004).

\bibitem{Ale96} I. L. Aleiner and A. I. Larkin, Phys. Rev. B {\bf 54}, 14423 (1996).

\bibitem{Ada03} I. Adagideli, Phys. Rev. B {\bf 68}, 233308 (2003). 



%\bibitem{Alt85} B. L. Altshuler, JETP Lett. {\bf 41}, 648 (1985).
%\bibitem{Lee85} P. A. Lee and A. D. Stone, Phys.\ Rev.\ Lett.\ {\bf 55}, 1622 (1985).
%\bibitem{Bar94} H. U. Baranger and P. A. Mello,  Phys.\ Rev.\ Lett.\ {\bf 73}, 142 (1994).
%\bibitem{Jal94} R. A. Jalabert, J.-L. Pichard, and C. W. J. Beenakker, Europhys.\ Lett.\ {\bf 27}, 255 (1994).
%\bibitem{Vav02} M. G. Vavilov and A. I. Larkin, Phys.\ Rev.\ B {\bf 67}, 115335 (2003).
%\bibitem{Sil03} P. G. Silvestrov, M. C. Goorden, and C. W. J. Beenakker, Phys.\ Rev.\ B {\bf 67}, 241301 (2003).
%\bibitem{Lod98} A. Lodder and Yu. V. Nazarov, Phys. Rev. B {\bf{58}}, 5783 (1998).
%\bibitem{Sil03L} P. G. Silvestrov, M. C. Goorden, and C. W. J. Beenakker, Phys.\ Rev.\ Lett.\ {\bf 90}, 116801 (2003).
%\bibitem{Aga00} O. Agam, I. Aleiner, and A. Larkin, Phys.\ Rev.\ Lett.\ {\bf 85}, 3153 (2000).
%\bibitem{Gor03}  M.C. Goorden, Ph. Jacquod, and C.W.J. Beenakker, Phys. Rev. B {\bf 68}, 220501 (2003).  
%\bibitem{Bog92} E. B. Bogomolny, Nonlinearity\ {\bf 5}, 805 (1992).
%\bibitem{Pra03} R. E. Prange,  Phys.\ Rev.\ Lett.\ {\bf 90}, 070401 (2003).
%\bibitem{Bor91} F. Borgonovi, I. Guarneri, and D. L. Shepelyansky, Phys.\ Rev.\ A {\bf 43}, 4517 (1991).
%\bibitem{Bor92} F. Borgonovi and I. Guarneri, J.\ Phys.\ A {\bf 25}, 3239 (1992).
%\bibitem{Fyo00} Y. V. Fyodorov and H.-J. Sommers, JETP Lett.\ {\bf 72}, 422 (2000).
%\bibitem{Oss02} A. Ossipov, T. Kottos, and T. Geisel, Europhys.\ Lett.\ {\bf 62}, 719 (2003).
%\bibitem{Kor03} A. Kormanyos, Z. Kaufmann, C. J. Lambert, and J. Cserti, Phys. Rev. B {\bf 67}, 172506 (2003).
%\bibitem{Jac04} Ph. Jacquod and E. V. Sukhorukov, cond-mat/0311528.

\end{thebibliography}
\end{document}